\documentclass[twocolumn,trackchanges]{aastex63}
\usepackage{amsmath}
\usepackage{float}
\usepackage{url}
\usepackage{graphicx}
\usepackage{threeparttable} 
\usepackage{multirow}
\usepackage[figuresright]{rotating}

\shorttitle{New Symbiotic stars}
\shortauthors{Chen et al.}

\graphicspath{{./}{figures/}}

\begin{document}

\title{New Symbiotic Stars from LAMOST DR10 Spectra and Multi-band Photometry}

\correspondingauthor{Liang Wang, Jing Chen} 
\email{liangwang@niaot.ac.cn, chenjing@niaot.ac.cn}


\author[0000-0001-8869-653X]{Jing Chen}
\affiliation{Nanjing Institute of Astronomical Optics \textnormal{\&} Technology, Chinese Academy of Sciences, Nanjing 210042, China}
\affiliation{CAS Key Laboratory of Astronomical Optical \textnormal{\&} Technology, Nanjing Institute of Astronomical Optics \textnormal{\&} Technology, Nanjing 210042, China}

\author[0000-0003-3603-1901]{Liang Wang}
\affiliation{Nanjing Institute of Astronomical Optics \textnormal{\&} Technology, Chinese Academy of Sciences, Nanjing 210042, China}
\affiliation{CAS Key Laboratory of Astronomical Optical \textnormal{\&} Technology, Nanjing Institute of Astronomical Optics \textnormal{\&} Technology, Nanjing 210042, China}
\affiliation{University of Chinese Academy of Sciences, Beijing 100049, China}

\author[0000-0001-7607-2666]{Yin-Bi Li}
\affiliation{CAS Key Laboratory of Optical Astronomy, National Astronomical Observatories, Beijing 100101, China}

\author[0000-0002-9279-2783]{Xiao-Xiao Ma}
\affiliation{University of Chinese Academy of Sciences, Beijing 100049, China}
\affiliation{CAS Key Laboratory of Optical Astronomy, National Astronomical Observatories, Beijing 100101, China}

\author[0000-0001-7865-2648]{A-Li Luo}
\affiliation{University of Chinese Academy of Sciences, Beijing 100049, China}
\affiliation{University of Chinese Academy of Sciences,Nanjing 211135,China}
\affiliation{CAS Key Laboratory of Optical Astronomy, National Astronomical Observatories, Beijing 100101, China}

\author{Zi-Chong Zhang}
\affiliation{Nanjing Institute of Astronomical Optics \textnormal{\&} Technology, Chinese Academy of Sciences, Nanjing 210042, China}
\affiliation{CAS Key Laboratory of Astronomical Optical \textnormal{\&} Technology, Nanjing Institute of Astronomical Optics \textnormal{\&} Technology, Nanjing 210042, China}
\affiliation{University of Chinese Academy of Sciences, Beijing 100049, China}

\author{Ming-Yi Ding}
\affiliation{Nanjing Institute of Astronomical Optics \textnormal{\&} Technology, Chinese Academy of Sciences, Nanjing 210042, China}
\affiliation{CAS Key Laboratory of Astronomical Optical \textnormal{\&} Technology, Nanjing Institute of Astronomical Optics \textnormal{\&} Technology, Nanjing 210042, China}

\author[0000-0001-9788-6378]{Kai Zhang}
\affiliation{Nanjing Institute of Astronomical Optics \textnormal{\&} Technology, Chinese Academy of Sciences, Nanjing 210042, China}
\affiliation{CAS Key Laboratory of Astronomical Optical \textnormal{\&} Technology, Nanjing Institute of Astronomical Optics \textnormal{\&} Technology, Nanjing 210042, China}
\affiliation{University of Chinese Academy of Sciences, Beijing 100049, China}

\begin{abstract}

Symbiotic star (SySt) is long-period interacting binary system, typically consisting of a white dwarf and a red giant surrounded by a nebula. These systems are natural astrophysical laboratories for investigating binary star evolution. In this paper, we identified nine SySts from the LAMOST DR10 low-resolution spectra survey, seven of which were previously known, while two are newly identified. Initially, we selected LAMOST spectra exhibiting typical SySt emission lines (e.g., $\rm H_{\alpha}, ~H_{\beta}, ~H_{\gamma}, ~and ~He II$). Subsequently, we utilized the distribution of known SySts on the HR diagram to select SySt candidates, and visually inspected their spectra. Ultimately, we classified all nine as S-type SySts using the $J - H$ vs. $H - K$ diagram. Additionally, based on multi-band photometric data from GALEX, Gaia, 2MASS, ALLWISE, and several X-ray catalogs, we found 12 accreting-only SySt (acc-SySt) candidates, characterized by concurrent ultraviolet and infrared excess and accretion process. Furthermore, we estimated the white dwarf temperatures by fitting their observed SEDs using a combination of Kurucz stellar atmosphere model and Koester white dwarf model. We compared the accretion rates of acc-SySt candidates and confirmed SySts, and found they have similar accretion rate distribution, providing evidence that these acc-SySt candidates constitute bona fide SySts.

\end{abstract}


\section{Introduction} \label{sec:intro}

The definition and understanding of symbiotic stars (SySts) have developed progressively over several decades. It began with the compilation of the HD catalog \citep{1912AnHar..56..165F, 1932PASP...44...56M}, where the spectra of several stars, such as Z And \citep{1911HarCi.168....1C, 1931PDAO....4..119P} and AG Peg \citep{1941PASP...53..124M}, exhibited remarkably strong high-ionization emission lines superimposed on the absorption spectra of red giants, accompanied by periodic photometric variations. As more stars with similar spectral characteristics were discovered, \cite{1942ApJ....95..386M} suggested use SySts to describe such objects. \cite{1932PASP...44..318B} and \cite{1936PAAS....8...14H} suggested that the peculiar combination spectra of SySts can be interpreted by a binary system, which was later championed by \cite{1940ApJ....91..546S, 1941ApJ....94..291S}. After decades of efforts, visual and ultraviolet spectroscopy finally showed conclusively that SySts are long-period interacting binary consisting of three components: a hot compact object, such as a white dwarf (WD), neutron star, or main sequence star with an accretion disk; a cool giant star, typically a red giant branch (RGB) or an asymptotic giant branch star; and an ionized nebula, which forms from the material expelled by the cool giant companion \citep{1966SvA....10..331B,1969IzKry..39..124B, 1986syst.book.....K, 2006ApJ...641..479H, 2019arXiv190901389M}. In many SySts, the cool giant companion loses mass via stellar winds or Roche-lobe overflow, while the hot companion accretes sufficient material to trigger a variety of physical phenomenon that can offer valuable insights into a number of important astrophysical topics \citep{2007BaltA..16....1M}, such as mass loss in evolved stars, the accretion mechanism of stellar winds, accretion disc formation, and formation of jets \citep{2003ASPC..303..376T, 2017MNRAS.468.4465C}. SySts are also considered as progenitors of Ia supernovae and serve as significant sources of both soft and hard X-rays emissions\citep{1992ApJ...397L..87M, 2016MNRAS.461L...1M}.

The SySts are classified into two main categories based on their near-infrared colors, i.e., ``stellar" (S) type and ``dusty" (D) type  \citep{1975MNRAS.171..171W}. S-type SySts exhibit typical colors of RGB stars, while the colors of D-type SySts indicate a significant contribution from a warm, dusty circumstellar envelope surrounding more evolved asymptotic giant branch stars \citep{2019ApJS..240...21A}. S-type SySts comprise approximately 70-80\% of the known SySts population \citep{2008AA...480..409C, 2019ApJS..240...21A}. SySts are usually identified through their optical spectra, which display both absorption features from the cool companion's photosphere (e.g., TiO, VO, $\rm C_2$ and CN bands) and strong emission lines produced by a luminous and hot WD(e.g., $\mathrm{HeII~\lambda}$4686\AA, $\mathrm{O VI~\lambda}$6830\AA, $\mathrm{[O III]~\lambda \lambda}$4959\AA, 5007\AA ~and prominent Balmer lines). 

Systematic searches for SySts have been conducted in recent years by various surveys, resulting in the discovery of approximately 400 SySts to date. Using the 3.9m Anglo-Australian Telescope, \cite{1984PASA....5..369A} identified 129 StSts and 15 SySt candidates. \cite{2008AA...480..409C} found 1183 candidates from the INT Photometric H-Alpha Survey (IPHAS), seven of which were later confirmed as genuine SySts through follow-up spectroscopic observation. \cite{2023MNRAS.519.6044A} utilized data from GALEX, 2MASS and AllWISE photometric surveys, and found 814 SySt candidates, and identified two genuine SySts in subsequent spectroscopic observations. The most extensive catalog of SySts to date comes from \cite{2019ApJS..240...21A}, who summarized a list of 323 known SySts, 257 of which are located in the Milky Way, while 66 are extragalactic. Although the number of known SySts has steadily increased over years, the total number of confirmed SySts still remains significantly lower compared to the expected number of SySts in our Galaxy, which range from $3 \times 10^3$ to $4 \times 10^5$ \citep{1984PASA....5..369A, 2003ASPC..303..539M}. This suggests that a large number of SySts remain undetected, awaiting future discovery.

It is noteworthy that the SySts identified so far have primarily been recognized based on the presence of strong emission lines in their optical spectra. However, many SySts have been identified with UV or X-ray observations, for example, R Aqr \citep{1980ApJ...237..506M} was identified based on UV emission and V934 Her \citep{1999AA...347..473G} was identified as a symbiotic X-ray binaries, consisting of a neutron star accreting mass from an M giant \citep{2019ApJ...872...43H}. \cite{2016MNRAS.461L...1M} identified SU Lyn as a SySt, although it features a typical M-type giant with a weak $\rm H_{\alpha}$ emission line and lacks other typical emission lines such as $\rm HeII, ~[Fe VII], ~ and~ [O III]$ \citep{ 2021MNRAS.505.6121M}. This absence of emission lines is hypothesized to be a consequence of a low accretion rate, which suppresses both stable hydrogen burning on the white dwarf's surface and the ionization of the cool giant's stellar wind \citep{2024ApJ...962..126X}. \cite{2016MNRAS.461L...1M} suggested that a significant number of SySts with low accretion rates (known as accreting-only SySt, i.e., acc-SySt), may have been overlooked in previous low-resolution spectroscopy surveys. This hypothesis has sparked some studies aimed at discovering acc-SySts. For instance, \cite{2021MNRAS.505.6121M} discovered 33 acc-SySt candidates through the GALactic Archaeology with HERMES (GALAH) high-resolution spectroscopic survey of the Southern hemisphere. \cite{2022AA...661A.124M} identified a new acc-SySts with VLT/X-shooter and Swift/UVOT data. \cite{2024ApJ...962..126X} found 10 acc-SySt candidates by cross-correlating Gaia, GALEX, and XMM-Newton catalogs. Despite the existence of acc-SySts, the current number of known SySts remains significantly lower than theoretical predictions, highlighting the need for further searches.

The Large Sky Area Multi-Object Fiber Spectroscopic Telescope (LAMOST), as a large-scale spectroscopic survey with tens of millions of spectra, provides low-resolution spectra (LRS) with a wavelength range from 3700 $\mathrm{\AA}$ to 9000 $\mathrm{\AA}$, which cover the important emission line characteristics of SySts. This renders LAMOST highly suitable for detecting SySts. \cite{2015RAA....15.1332L} discovered two SySts using the first three years of LAMOST spectral data, although one was later confirmed by \cite{2020CoSka..50..672A} not to be a SySt. \cite{2023RAA....23j5012J} applied machine learning algorithm combined with 2MASS and WISE photometric data to identify two SySts and 11 acc-SySt candidates from LAMOST DR10. 

In this paper, we conducted a systematic search for SySts and acc-SySt candidates using the LAMOST LRS and GALEX photometry, respectively. The paper is organized as follows: Section \ref{sec:data} describes the primary survey data utilized in this study. The methodology for identifying SySts from the LAMOST survey is detailed in Section \ref{sec:syst}. Section \ref{sec:accrete} outlines the process of selecting acc-SySt candidates. In Section \ref{sec:discuss}, we analyze the white dwarf temperatures and accretion rates of the acc-SySt candidates. Finally, the summary is presented in Section \ref{sec:summary}.

\section{Data} \label{sec:data}

\subsection{LAMOST}

LAMOST is a 4 m Schmidt telescope with a wide view field of 20 $\mathrm{deg^2}$ in the sky, and equipped with 4000 fibers \citep{2006ChJAA...6..265Z, 2012RAA....12.1243L}. Since 2012 September, LAMOST officially conducted its regular survey and had released its tenth dataset (DR10)\footnote{\href{lamostdr10}{http://www.lamost.org/dr10/
}} version 1.0 by March 2023. In this data release, there are totally 11,817,430 low resolution spectra, including 11,473,644 stellar spectra, 263,444 galaxy spectra, and 80,342 quasar spectra, and they cover the wavelength range of 3700 $\mathrm{\AA}$ -- 9000 $\mathrm{\AA}$ with a spectral resolution of 1800 (R $\sim$ 1800) at 5500 $\mathrm{\AA}$. Furthermore, this data release also publish 11 spectroscopic parameter catalogs, and the LAMOST LRS General Catalog was utilized in this work.

\subsection{Gaia}

Gaia, the European Space Agency's space-astrometry mission, is primarily designed to investigate the kinematic, dynamical, chemical structure and evolution of our Milky Way \citep{2016AA...595A...1G}. The mission aims to build the most comprehensive and precise three-dimensional map of our galaxy. The third data release of Gaia provided astrometric measurements and broad-band photometry in the $G$, $ G_{BP}$, and $G_{RP}$ for approximately 1.8 billion objects, based on 34 months of observations \citep{2023AA...674A...1G}. The magnitudes, colors, and parallaxes from the Gaia DR3 main source catalog were utilized to constrain the selecting process for identifying SySts.

\subsection{GALEX}

The Galaxy Evolution Explorer (GALEX) survey \citep{2005ApJ...619L...1M} imaged the sky with two ultraviolet (UV) bands. The far-UV band (FUV), with an effective wavelength of $\mathrm{\lambda_{eff} \sim 1258 \AA}$, spans the spectral range from 1340 $\mathrm{\AA}$ to 1806 $\mathrm{\AA}$, while the near-UV band (NUV), with an effective wavelength of $\mathrm{\lambda_{eff} \sim 2310\AA}$, covers 1693 $\mathrm{\AA}$ to 3007 $\mathrm{\AA}$. The GALEX GR6/7 AIS source catalog includes 82,992,086 unique UV sources \citep{2017ApJS..230...24B}. We aimed to use GALEX data to search for WDs, as GALEX has been proven effective in identifying new WDs, either isolated or in binary systems \citep{2011MNRAS.411.2770B}.

\section{Search for Symbiotic stars} \label{sec:syst}

We searched for SySts by examining the characteristics of emission lines in the optical spectra, following a series of steps: First, we selected spectra with emission lines from the entire dataset of LAMOST DR10 LRS; Next, we identified SySt candidates from these emission-line stars based on the established distribution of known SySts in the Hertzsprung-Russell diagram (HRD); Finally, we conducted a visual inspection of the identified SySt candidates spectra to confirm the presence of SySts.

\subsection{Selection of emission-line stars}

Before identifying emission-line stars, we needed to preprocess the LAMOST one-dimensional LRS. The preprocessing steps were as follows: (1) The pixels with `ormask' = 0 in each spectrum were selected to ensure that no bad conditions occured in any exposure. (2) The `z' value in each spectrum, representing the redshift of the target and primarily determined by the LAMOST spectroscopy pipeline, was taken into account, and the spectra were shifted to align all spectral lines with their rest-frame laboratory wavelengths. (3) To address the missing values at wavelengths below 3800 $\mathrm{\AA}$, we established a lower wavelength limit of 3800 $\mathrm{\AA}$. (4) Each spectrum was resampled to 1 $\mathrm{\AA}$. (5) Finally, we normalized each spectrum using min-max normalization.

\begin{figure*}[ht!] 
	\centering
        \includegraphics[width=1.0\textwidth]{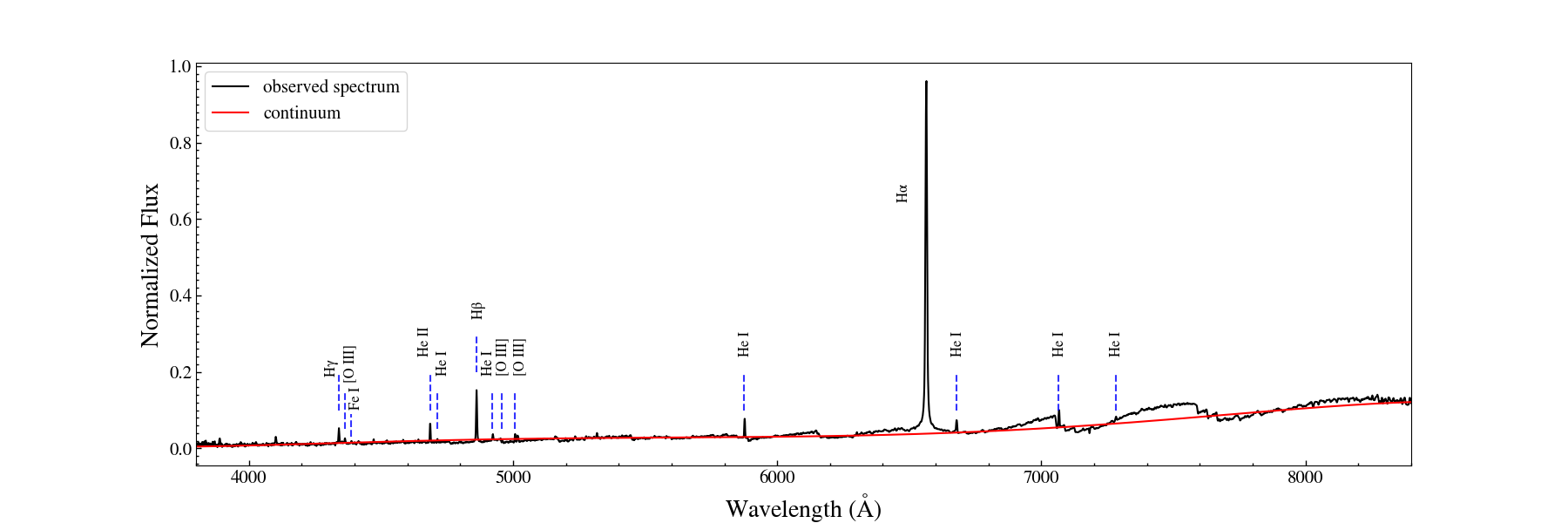}
	\caption{LAMOST LRS of V2428 Cyg. The black and red lines are the observed spectrum and pseudo-continuum of the objects, respectively.\label{Continua}}
\end{figure*}

The preprocessed spectra were fitted using a fifth-order polynomial to select strong lines. These strong lines were subsequently masked, and the remaining spectra were refitted with a fifth-order polynomial to construct the pseudo-continuum, as in Fig.\ref{Continua}. Given that some emission lines in SySts are weak, points that are only 0.5$\sigma$ above the pseudo-continuum were marked as potential emission lines, where $\sigma$ is the standard deviation of the pseudo-continuum. To mitigate the risk of missing spectra due to weak emission lines not being detectable under the criteria, we focused specifically on a few strong emission lines, including $\mathrm{H_{\alpha}}$, $\mathrm{H_{\beta}}$, $\mathrm{ H_{\gamma}}$, and $\mathrm{HeII}$. Spectra exhibiting these strong emission lines were classified as emission-line stars. Ultimately, we identified 58,110 spectra containing strong emission lines from the LAMOST DR10 LRS dataset.

\subsection{Selection of Symbiotic stars} \label{sec:select_sybt}

\begin{figure}[ht!] 
	\centering
        \includegraphics[width=0.45\textwidth]{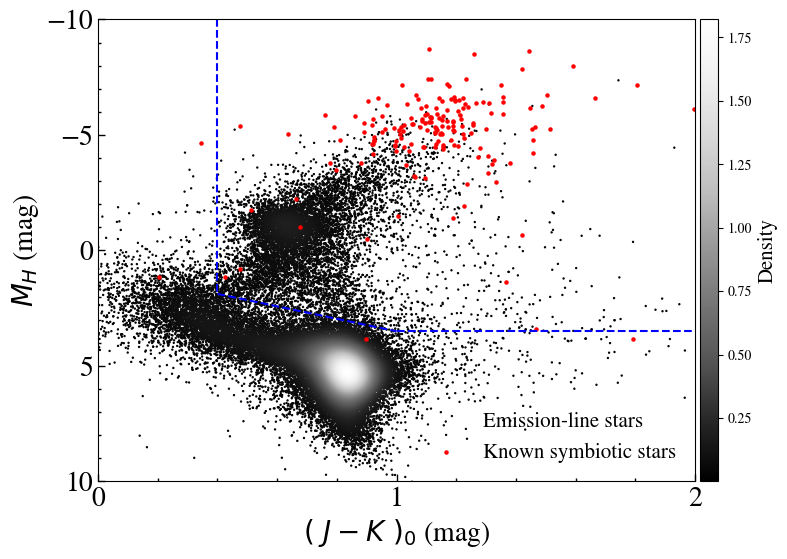}
	\caption{The positions of the selected emission-line stars in the Hertzsprung-Russell diagram, with a color-coded density distribution indicating their distribution. The red dots denote the known symbiotic stars as cataloged by \cite{2019ApJS..240...21A}. The blue dashed line delineates the criteria used for selecting the symbiotic candidates. \label{CMD}}
\end{figure}

To identify SySts from the emission-line stars, we initially selected SySt candidates based on the distribution of known SySts in the HRD, which is known to come from one of the largest SySts catalogs compiled by \cite{2019ApJS..240...21A}, indicated by the red dots in Fig.\ref{CMD}. It is evident that most SySts are giants, though some are found within the main sequence region. The black dots in Fig.\ref{CMD} represent the positions of the selected emission-line stars in the HRD. 
The distances used for absolute magnitude and extinction calculations were not derived from the simple inversion of Gaia parallaxes, as this naive approach becomes unreliable beyond 2 kpc where Gaia parallax uncertainties exceed 5\% \citep{2018AA...616A...9L}. Instead, we adopted the Bayesian distance estimates from \cite{2018AJ....156...58B}, which provide physically meaningful distances even for sources with negative parallax measurements or limited parallax precision. To maximize the number of potential SySt candidates, we adopted the RGB region defined by \cite{2022AJ....164..126A} and applied the following criterion $\rm M_H < 3.5, (J-K)_0 > 0.4, ~and ~M_H < 2.7(J-K)_0 + 0.8$, resulting in a total of 8,845 spectra being retained.

\begin{deluxetable*}{cccccccccc}
\tablecaption{The information of 9 symbiotic stars from LAMOST DR10. \label{Syst_Inf}}
\tabletypesize{\scriptsize}
\setlength\tabcolsep{2pt}
\tablehead{\colhead{Name} & \colhead{RA(J2000.0)} & \colhead{Dec(J2000.0)} & \colhead{T$eff$} & \colhead{log$g$} & \colhead{[Fe/H]} & \colhead{$\mathrm{type\_{L}}$} & \colhead{$\mathrm{type\_{S}}$} & \colhead{$V$} & \colhead{$Ks$}\\ 
\colhead{} & \colhead{(h:m:s)} & \colhead{(d:m:s)} & \colhead{(K)} & \colhead{(dex)} & \colhead{(dex)} & \colhead{} & \colhead{} & \colhead{(mag)} & \colhead{(mag)} } 
\startdata
LAMOST J122804.90-014825.7 & 12 28 04.9 & -01 48 25.6 & 4170.73 $\pm$ 43.24 & 1.25 $\pm$ 0.06 & -1.05 $\pm$ 0.03 & K & Symbiotic* & 12.64 $\pm$ 0.06  & 9.03 $\pm$ 0.02\\
AS  297 & 18 14 34.2 & +20 59 21.2 & 3293.17 $\pm$ 125.91 & 3.19 $\pm$ 0.37 & - & gM & Symbiotic* & 11.10 $\pm$ -  & 7.86 $\pm$ 0.02 \\
HD 342007 & 18 22 07.9 & +23 27 19.9 & 3013.42 $\pm$ 109.78 & 2.65 $\pm$ 0.42 & - & gM & Symbiotic*  & 11.52 $\pm$ 0.08 & 5.32 $\pm$ 0.02\\
IPHAS J184446.08+060703.5 & 18 44 46.1 & +06 07 03.6 & 4040.12 $\pm$ 109.78 & 2.52 $\pm$ 0.42 & - & gM & Symbiotic* & - & 9.41 $\pm$ 0.02\\
StHA  169 & 19 49 57.6 & +46 15 20.5 & 3622.77 $\pm$ 95.10 & 2.98 $\pm$ 0.25 & - & gM & Symbiotic* & 13.56 $\pm$ 0.13 & 8.93 $\pm$ 0.02\\
V2428 Cyg & 20 41 19.0 & +34 44 52.3 & 3318.99 $\pm$ 131.91 & 3.50 $\pm$ 0.46 & - & gM & Symbiotic* & 14.76 $\pm$ 0.01 & 7.95 $\pm$ 0.03\\
V1329 Cyg & 20 51 01.2 & +35 34 54.1 & 3162.27 $\pm$ 109.78 & 2.69 $\pm$ 0.42 & - & gM & Symbiotic* & 13.12 $\pm$ 0.01 & 6.75 $\pm$ 0.02\\
LAMOST J072528.17+342530.4 & 07 25 28.2 & +34 25 30.4 & 4315.04 $\pm$ 66.87 & 0.52 $\pm$ 0.11 & -2.32 $\pm$ 0.07 & K & Star & - & -\\
V758 Cyg & 20 00 23.0 & +44 23 59.2 & 3445.70 $\pm$ 109.78 & 3.27 $\pm$ 0.42 & - & gM & Mira & 12.96 $\pm$ - & 8.16 $\pm$ 0.02\\
\enddata
\tablecomments{`-' denotes the absence of a corresponding measurement. `$\mathrm{type\_{L}}$' and `$\mathrm{type\_{S}}$' represent spectral type from LAMOST pipeline and SIMBAD database, respectively.}
\end{deluxetable*}

Subsequently, we conducted a visual inspection of the 8,845 spectra focused on their strong emission lines, i.e., $\mathrm{H_{\alpha}}$ (6563 \AA), $\mathrm{H_{\beta}}$ (4861 \AA), $\mathrm{ H_{\gamma}}$ (4340 \AA), $\mathrm{HeII}$ (4686 \AA), $\mathrm{HeI}$ (4713 \AA, 5875 \AA, 6678 \AA) and $\mathrm{[O III]}$ (4363 \AA), as well as the TiO and VO molecular bands. During the visual inspection process, we found that low-temperature dwarf stars with chromospheric activity are the most easily confused with SySts. These stars exhibit spectral features characterized by both molecular bands and emission lines. To distinguish them, we excluded such objects based on the subclasses provided by the LAMOST pipeline and the existence of Ca II H \& K in their spectra. We identified nine SySts, which is detailed in Table \ref{Syst_Inf}. Columns 4 to 8 of Table \ref{Syst_Inf} present the effective temperatures, surface gravities, metallicities, along with their corresponding uncertainties, as well as classifications from the LAMOST stellar parameter pipeline. Notably, all identified stars are classified as low-temperature giants.
We cross-checked these stars with SIMBAD database, and their types are also listed in Table \ref{Syst_Inf}. We found seven of these stars were previously recognized as SySts \citep{2015RAA....15.1332L, 2010AA...509A..41C, 2019ApJS..240...21A}, while for rest two, V758 Cyg has been classified as a Mira variable, LAMOST J072528.17+342530.4 without any classification. We identified these two stars as newly SySts, based on their strong emission lines exhibited in LAMOST LRS.

\begin{figure*}[ht!] 
	\centering
        \includegraphics[width=1.0\textwidth]{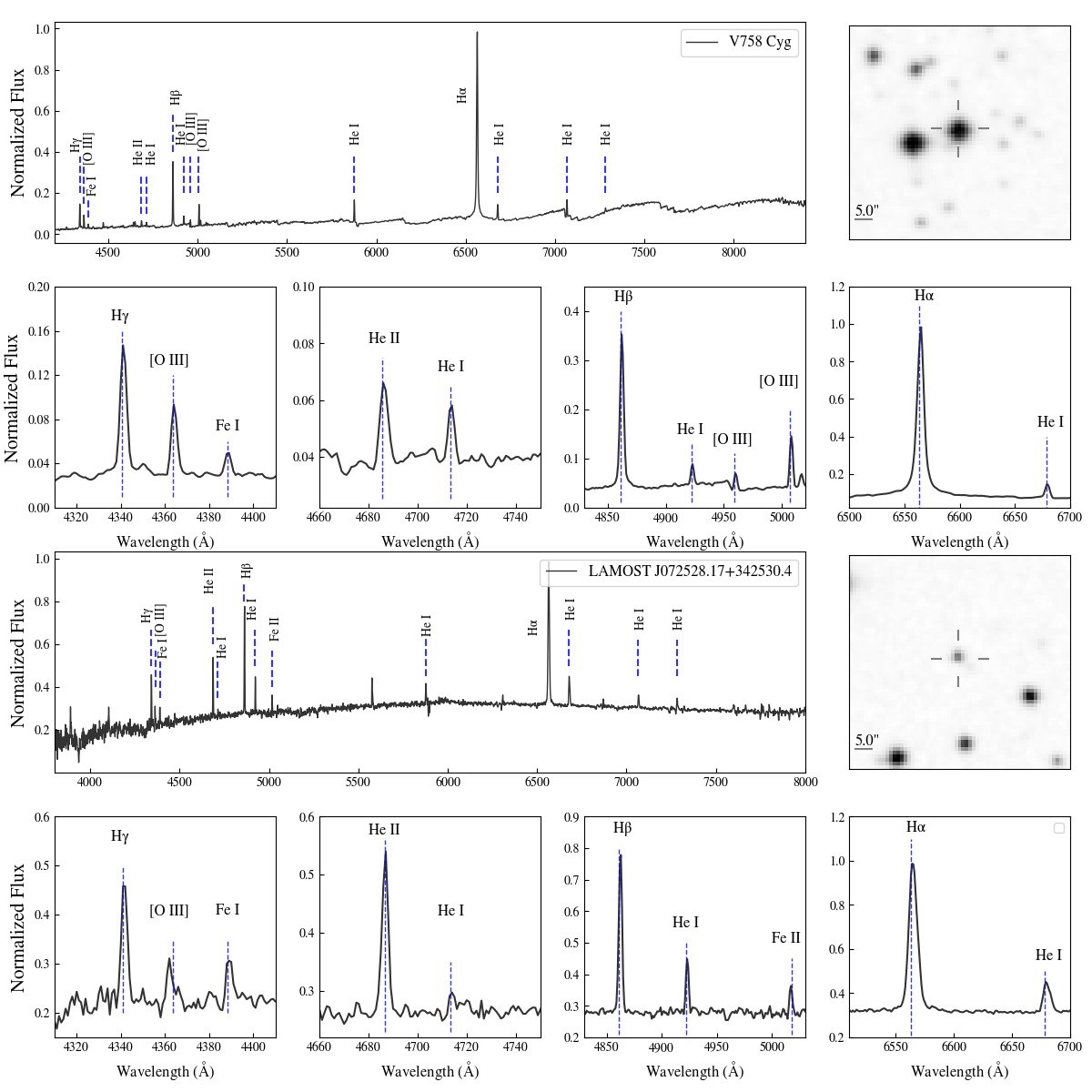}
	\caption{ The LAMOST low-resolution spectra of the two newly identified symbiotic stars are presented. The top left panel displays the observed spectrum, while the upper right panel shows the DSS DR2 red image of the source, with north oriented upwards and east to the left. The bottom panel provides a closer view of the $\mathrm{H_{\gamma}}$, [O III] 4363 \AA, He II 4686 \AA, He I 4713 \AA, $\mathrm{H_{\beta}}$, He I 4922 \AA, $\mathrm{H_{\alpha}}$ and He I 6678 \AA ~emission lines.
 \label{SySt}}
\end{figure*}

Figure \ref{SySt} displays the LAMOST LRS and Digitized Sky Survey (DSS) DR2 images with 60$"$ field of view of the two newly discovered SySts. A brief description of the characteristics of these two objects is provided below.
\begin{itemize}
    \item V758 Cyg. It has been classified as a Mira variable by \cite{2012AA...548A..79A} and a long period variable star by \cite{2022yCat.1358....0G}. However, in our study, we classify it as a SySt due to its characteristic spectrum, which includes prominent deep TiO absorption bands commonly observed in M-giants, as well as emission lines from low to high excitation, including: the H I Balmer series down to $\mathrm{H_{\gamma}}$; He I 5875, 6678 and 7065 $\mathrm{\AA}$; He II 4686 $\mathrm{\AA}$; [O III] 4363 4959, 5007 $\mathrm{\AA}$ and Fe I 4388 $\mathrm{\AA}$. From the DSS image, it can be seen that there is a star of similar brightness located 13$"$ to the lower left of V758 Cyg. According to Gaia DR3 data, this star is a foreground star of V758 Cyg, and has been classified by LAMOST as an F5 star, with no emission lines detected in its spectrum. Therefore, the emission lines observed in V758 Cyg do not contaminate from this foreground star.
    \item LAMOST J072528.17+342530.4. Its spectrum closely resembles with LAMOST J122804.90-014825.7, which was classified as a S-type SySt by \cite{2015RAA....15.1332L}. The strong He II 4686 $\mathrm{\AA}$ emission suggests the presence of a hot component within the object. Additionally, after cross-matching this target with the GALEX GR6/7 AIS source catalog, we found that $m_{\mathrm{FUV}} - m_{\mathrm{NUV}} < 1$, further supporting the existence of a hot component \citep{2007ApJS..173..659B}. Moreover, the LAMOST pipeline classifies it as type K3 and log $g=0.83$, indicating that it is a cool giant. Given these factors, along with the presence of characteristic emission lines typical of SySts, we identified it as a SySt.
\end{itemize}

\subsection{Classification of Symbiotic stars} \label{classification}

\begin{figure}[ht!] 
	\centering
        \includegraphics[width=0.45\textwidth]{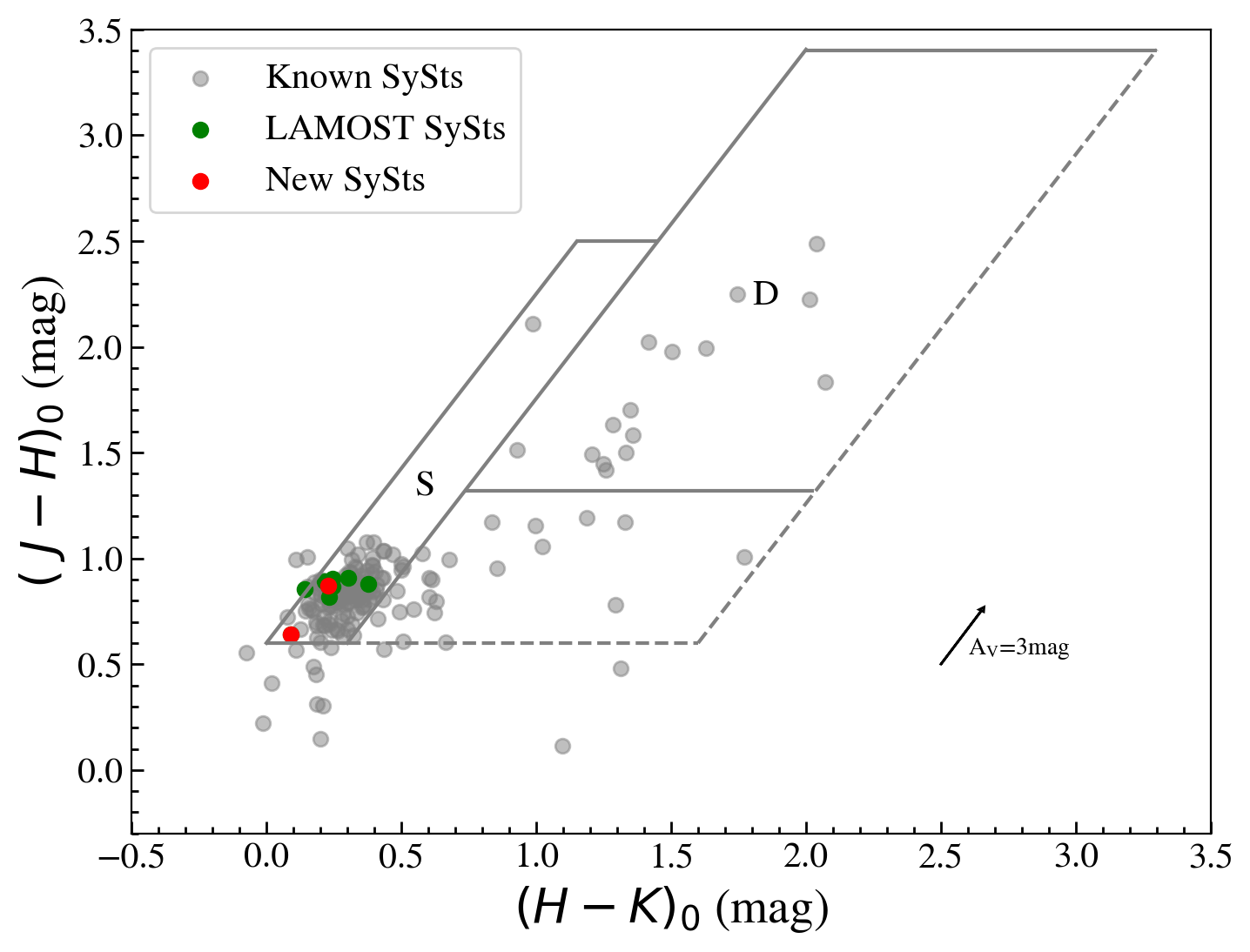}
	\caption{The $J$ - $H$ vs. $H$ - $K$ color-color diagram of the known SySts sample from \cite{2019ApJS..240...21A} (gray) and SySts from LAMOST DR10 LRS (green), and the two new identified SySts (red). The solid and dashed boxes indicate the region of S- and D-type SySts region defined in \cite{2008AA...480..409C} and \cite{2014AA...567A..49R}. \label{ccd}}
\end{figure}

The $J - H$ versus $H - Ks$ diagram of 2MASS \citep{2006AJ....131.1163S} has been extensively utilized to investigate the near-infrared properties of SySts, facilitating their classification into S-type and D-type, as well as the identification of new candidates \citep{2008AA...480..409C, 2014AA...567A..49R, 2019MNRAS.483.5077A}. We plotted the 2MASS colors of the nine SySts identified from LAMOST in Fig.\ref{ccd}, together with the known SySts samples from \cite{2019ApJS..240...21A}. 
The extinction in the $J$, $H$, and $Ks$ magnitudes was estimated using the Python package $dustmaps$ \footnote{\href{dustmaps}{https://pypi.org/project/dustmaps}} with three-dimensional Bayestar19 version \citep{2019ApJ...887...93G}, which models dust reddening based on the angular position and distance on the sky. Among the nine SySts observed with LAMOST, the two newly discovered SySts are highlighted in green, while the seven previously known SySts are denoted as red. Both groups fall within the region of S-type SySts, leading us to classify the two new SySts as S-type.

\section{Search for accreting-only Symbiotic candidates} \label{sec:accrete}

The acc-SySts exhibit either negligible or weak emission lines in their optical spectra. We identified these systems based on multi-wavelength photometric data from Gaia, 2MASS, ALLWISE, GALEX and several X-ray surveys, aiming to ensure the presence of both infrared and ultraviolet excesses, indicative of accretion processes.

\subsection{method}

Selecting acc-SySts with only photometric information would have some contaminates, such as stars exhibiting both ultraviolet and infrared excess, K/M giants with chromospheric activity, and other combinations involving ``WD + RG" like barium stars and Technetium-poor S stars, it is crucial to exclude these sources. According to \cite{2023MNRAS.519.6044A}, the criterion $m_{\mathrm{FUV}} - m_{\mathrm{NUV}} < 1$ effectively eliminates the vast majority of single K/M giants, barium stars and Technetium-poor S stars from the acc-SySt candidates. The preliminary sample includes sources that satisfy the $m_{\mathrm{FUV}} - m_{\mathrm{NUV}} < 1$ criterion from the GALEX GR6/7 AIS catalog, encompassing a total of 6,943,433 sources. This selection is intended to confirm the presence of a hot component, which may arise directly from the photospheric emission of the white dwarfs or indirectly from the surrounding accretion disc. The preliminary sample was then cross-matched with the 2MASS and ALLWISE catalogs in order to subsequently select candidates from the sample that exhibiting infrared excess. The cross-matching was performed using TOPCAT \citep{2005ASPC..347...29T} software with a 5$''$ matching radius, resulting in 1,093,194 common sources. 

We adopted the two different infrared excess criteria of S-type SySts as outlined by \cite{2019MNRAS.483.5077A, 2021MNRAS.502.2513A}, which are as follows:
\begin{enumerate}
    \item $J-H >= 0.78~\mathrm{and}~0 < K_s - W3 < 1.18~\mathrm{and}~ W1$ $- W2 < 0.09$,
    \item $J-H >= 0.78~\mathrm{and}~0 < K_s - W3 < 1.18~\mathrm{and}~ W1$ $ - W2 >= 0.09~\mathrm{and}~0 < W1 - W4 < 0.92$.
\end{enumerate}
The extinction for the $J$, $H$, and $Ks$ bands was estimated as described in Section \ref{classification}. The extinction for $W1$ through $W4$ was calculated using the average extinction law from \cite{2006ApJ...637..774C} via the Python package $dust\_extinction$ \footnote{\href{dustextinction}{https://dust-extinction.readthedocs.io/en/stable/index.html}}. The distance used for extinction calculations was from \cite{2018AJ....156...58B}, which was cross-matched with the 1,093,194 sources, yielding 344,324 common sources. These two criteria effectively distinguish S-type SySts from their mimics, achieving an accuracy of over 90\% for S-type SySts. The first criterion successfully identifies more than 80\% of S-type SySts, while the second criterion identifies approximately 10\%. We did not apply the infrared criteria for D-type SySts due to their broad distribution in the color-color diagrams, which could lead to contamination from various other stellar types. The more than 300,000 common sources that satisfy one of the above specified criteria have been retained as acc-SySt candidates. Among these, 369 sources meet criterion 1, and 38 meet criterion 2, resulting in a total of 407 sources.

\begin{figure}[ht!] 
	\centering
        \includegraphics[width=0.45\textwidth]{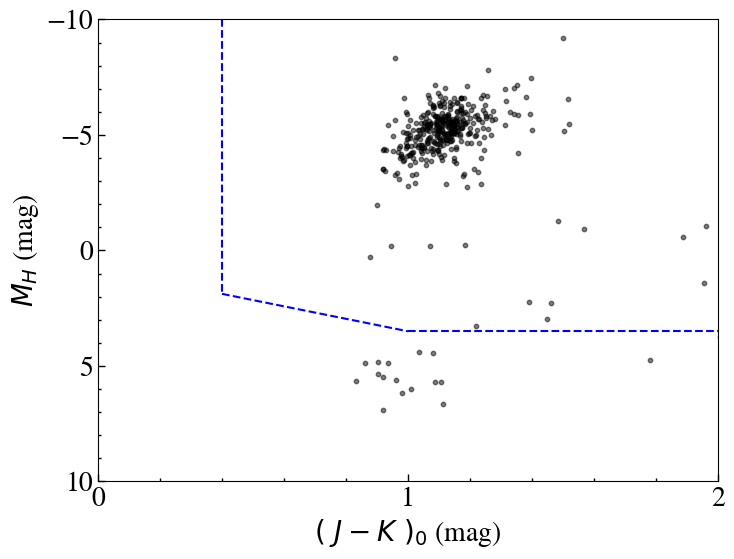}
	\caption{The 2MASS $\mathrm{M}_H~\mathrm{vs.}~(J-K)_0$ HR diagram for the 407 sources. The blue dashed lines represent the criteria in section \ref{sec:select_sybt}. \label{hrd}}
\end{figure}

While the incorporation of these bands in our selection criteria is essential, it may introduce both misclassifications and sample incompleteness. The photometric uncertainties in the 2MASS $J$, $H$, and $K$ bands are generally below 0.05 mag \citep{2022AA...662A.125F}. However, measurements in the $W3$ and $W4$ bands are significantly affected by contamination from interstellar dust emission, leading to elevated uncertainties \citep{2010AJ....140.1868W}. Additionally, although the typical photometric precision per epoch for GALEX is approximately 0.05 and 0.03 mag in the $FUV$ and $NUV$ bands, respectively; sources fainter than 21 mag in either band still suffer from substantial photometric errors \citep{2007ApJS..173..682M}. Therefore, additional verification and refinement of the selected candidates are necessary to ensure the reliability of the sample. Figure \ref{hrd} illustrates the $M_H~\mathrm{vs.}~(J-K)_0$ for the 407 sources, revealing two distinct regions: one corresponding to RGB and the other to the main sequence. The known SySts are predominantly found in the RGB phase, the main sequence phase may include a significant number of unknown contaminates and need to be further identification. To enhance the reliability of our sample, we further filtered the candidates to focus on those predominantly in the RGB evolutionary stage using the criteria in section \ref{sec:select_sybt}. This refinement resulted in a final selection of 387 sources.

\subsection{Cross-matching with SIMBAD}

To further enhance the purity of our acc-SySts candidates samples, we cross-matched the 387 sources with the SIMBAD database, and excluded contaminates that had previously been classified as K/M giants with chromospheric activity, barium stars and Technetium-poor S stars.

We cross-matched the 387 sources with SIMBAD and obtained 296 common sources. Previous studies indicate that the vast majority of the samples are unstudied stars and long-period variable stars. Eleven stars were identified as spectroscopic binaries in the Non-single stars with Single Lined Spectroscopic binary model catalogue by Gaia DR3 \citep{2022yCat.1357....0G}, we included these spectroscopic binary in the final sample. Additionally, we excluded 10 stars classified as S-type and 6 stars as Systs to maintain sample integrity. As carbon stars and chemically peculiar stars may contain barium stars, these were also omitted from our final sample. Ultimately, the final count of 270 stars left, along with the 91 stars that were not cross-matched with SIMBAD, total 361 stars reported as final sample.

\subsection{X-ray Counterparts}

\cite{2021MNRAS.505.6121M} noted that acc-SySts with low luminosities require robust evidence for accretion, such as flickering or X-ray emission, which rules out the presence of a non-accreting WD. To maximize the collection of X-ray fluxes for acc-SySt candidates, we compiled data from several X-ray survey catalogs, a point source catalog from all ROSAT PSPC pointed observations \citep{2000yCat.9031....0W}, the Chandra Source Catalog version 2.1 \citep{2010ApJS..189...37E}, the fourth XMM-Newton source catalogue (4XMM-DR13) \citep{2020AA...641A.136W}, an improved and expanded Swift X-ray Telescope Point Source catalog \citep{2020ApJS..247...54E}, the detailed information are summarized in Table \ref{tab:Xray_Catalog}.

Additionally, the first six months of the SRG/eROSITA all-sky survey data have been published \citep{2024AA...682A..34M}, and was used in this analysis. In Table \ref{tab:Xray_Catalog}, we lists the number of sources in each catalog, the time span over which data were collected, the energy coverage of each survey, and the number of common sources matched with the final sample. Since some sources are identified in multiple X-ray catalogs, we retained only those detections from catalogs with broader energy coverage. Following this selection, a total of 12 acc-SySt candidates have associated X-ray flux measurements, as summarized in Table \ref{tab:Accret_inf}. Columns 4 and 5 list the X-ray survey and the corresponding X-ray fluxes with their uncertainties. Column 6 provides the distances and associated errors. Columns 7 and 8 present the estimated accretion rates and white dwarf temperatures, which will be discussed in detail in Section \ref{sec:discuss}.

\begin{deluxetable*}{cccccc}[ht!]
\tablecaption{Cross-matching with X-ray catalogs. \label{tab:Xray_Catalog}}
\setlength\tabcolsep{10pt}
\tablehead{\colhead{Catalogue[Mission]} & \colhead{$\mathrm{N_{objects}}$} & \colhead{Time span} & Energy coverage & Common sources & Reference} 
\startdata
WGACAT [ROSAT] & 88,621 & 1991--1995 & 0.1--2.4 keV & 1 & (1) \\
CSC2.1 [Chandra] & 407,806 & 1999--2022 & 0.2--7.0 keV & 2 & (2) \\
4XMM-DR13 [XMM-Newton] & 983,948 & 2000--2022 & 0.2--12 keV & 4 & (3) \\
2SXPS [Swift] & 206,335 & 2005--2018 & 0.3--10 keV & 6 & (4) \\
eRASS1 Main [SRG/eROSITA] & 930,203 & 2019--2020 & 0.2--2.3 keV & 4 & (5) \\
eRASS1 Supp [SRG/eROSITA] & 347,274 & 2019--2020 & 0.2--2.3 keV & 1 & (5) \\
eRASS1 Hard [SRG/eROSITA] & 5,466 & 2019--2020 & 2.3--5.0 keV & 1 & (5) 
\enddata
\tablecomments{References. (1):\cite{2000yCat.9031....0W};(2):\href{https://cxc.cfa.harvard.edu/csc/}{https://cxc.cfa.harvard.edu/csc/};(3):\cite{2020AA...641A.136W}; (4):\cite{2020ApJS..247...54E}; (5):\cite{2024AA...682A..34M}}
\end{deluxetable*}

\begin{sidewaystable}[thp]
\caption{The information for the 12 acc-SySt candidates.}\label{tab:Accret_inf}
\setlength\tabcolsep{1.0pt}
\begin{tabular*}{\textheight}{@{\extracolsep\fill}ccccccccc}
\toprule%
Name & RA(J2000) & DEC(J2000) & Source of X-ray & $f_{X}$ & $d$ & $M_{acc}$ & $\mathrm{WD_{teff}}$ & $\mathrm{type\_S}$ \\ & (deg) & (deg) & & $\mathrm{(erg^{-1} cm^{-2} s^{-1})}$ & (pc) & $\mathrm{(M_{\odot}/yr)}$ & K &  \\ \hline
GALEX J032336.4-195301 & 50.901824 & -19.883730 & eRASS1 Main & (7.66±2.16)E-14 & 715.29±90.62 & (4.88±1.80)E-13 & 10500±125 & LongPeriodV* \\
GALEX J064255.1+552827 & 100.729893 & 55.474185 & 2SXPS & (4.93±0.33)E-12 & 658.83±42.67 & (2.66±0.40)E-11 & 13250±125 & LongPeriodV*\\
GALEX J073945.7+653908 & 114.940690 & 65.652349 & 4XMMDR13 & (3.56±2.92)E-14 & 2385.48±209.97 & (2.52±2.10)E-12 & 12750±125 & Star\\
GALEX J083531.3-090417 & 128.880659 & -9.071398 & 2SXPS & (1.18±0.12)E-12 & 3763.14±546.91 & (2.08±0.60)E-10 & 13000±125 & RotV*\\
GALEX J084621.2+013756 & 131.588369 & 1.632288 & 4XMMDR13 & (5.93±0.16)E-13 & 392.61±20.52 & (1.14±0.12)E-12 & - & Mira\\
GALEX J102820.3+683812 & 157.084706 & 68.636929 & 4XMMDR13 & (2.50±1.04)E-14 & 3607.60±477.69 & (4.05±2.00)E-12 & 10500±125 & LongPeriodV*\_Candidate\\
GALEX J105840.0-272959 & 164.666864 & -27.499873 & eRASS1 Supp & (3.56±1.88)E-14 & 2184.05±334.03 & (2.11±1.30)E-12 & 10500±125 & LongPeriodV*\_Candidate\\
GALEX J130937.0-272726 & 197.404320 & -27.457281 & eRASS1 Main & (5.20±2.15)E-14 & 1409.78±198.78 & (1.29±0.60)E-12 & 10000±125 & Star\\
GALEX J143518.4-132914 & 218.826952 & -13.487421 & eRASS1 Main & (1.04±0.30)E-13 & 1590.41±221.13 & (3.27±1.30)E-12 & 10250±125 & LongPeriodV*\_Candidate\\
GALEX J170634.5+235818 & 256.643923 & 23.971864 & 4XMMDR13 & (1.62±0.01)E-10 & 536.10±9.27 & (5.79±0.20)E-10 & 10500±125 & LowMassXBin \\
GALEX J180704.7+111533 & 271.769626 & 11.259364 & 2SXPS & (8.85±5.15)E-14 & 2492.12±725.74 & (6.84±6.00)E-12 & 10750±125 & LongPeriodV* \\
GALEX J211001.4+040235 & 317.506070 & 4.043300 & 2SXPS & (1.44±0.55)E-13 & 2761.51±458.46 & (1.37±0.70)E-11 & 12750±125 & Star\\
\botrule
\end{tabular*}
\end{sidewaystable}

\section{Analysis}\label{sec:discuss}

In this section, we examine the temperature of white dwarf of acc-SySt candidates and their corresponding accretion rates.

\subsection{The Temperature of White Dwarf}

The effective temperature of the white dwarf in acc-SySt candidates is a fundamental parameter that we can extract from the spectral energy distribution (SED) analysis. UV fluxes provide critical constraints on the physical parameters of the white dwarf, given that its emission typically dominates the SED in this wavelength regime. Due to the acc-SySt candidates compose a white dwarf and a red giant, to create the SEDs, we use the photometric bands from GALEX \citep{2017ApJS..230...24B}, Gaia DR3 \citep{2023AA...674A...1G}, 2MASS \citep{2006AJ....131.1163S}, and WISE \citep{2010AJ....140.1868W}. The E(B$-$V) value is given for each source in the GALEX catalog, and the distances were from \cite{2018AJ....156...58B}.

In the process of fitting SED, we use VOSA\footnote{\href{VOSA}{http://svo2.cab.inta-csic.es/theory/vosa/}} \citep{2008AA...492..277B}, developed by the Spanish Virtual Observatory, which can automatic fitting synthetic spectra or photometry extracted from theoretical models to observational data. VOSA contains several catalogs with observed photometry from the infrared to the ultraviolet and 70 collections of theoretical spectra and observational templates. We use the $binary ~fit$ of VOSA to fit the SED, and account for the red giants of the SED with a Kurucz stellar atmosphere model \citep{1997AA...318..841C} and white dwarf of the SED with the Koester WD models \citep{2010MmSAI..81..921K}. The Koester stellar atmosphere models span a wide effective temperature range from 5,000 K to 80,000 K. The sampling across the parameter space is non-uniform, and are computed in 250 K steps between 5,000 K and 20,000 K. As pointed out by \cite{2022AJ....164..126A}, because of the saturation of the (FUV$–$NUV) color, for WDs hotter than 30,000 K we cannot retrieve reliable temperatures from the UV GALEX bands; fortunately, the vast majority of WDs have temperatures lower than 30,000 K.

\begin{figure}[ht!] 
	\centering
        \includegraphics[width=0.45\textwidth]{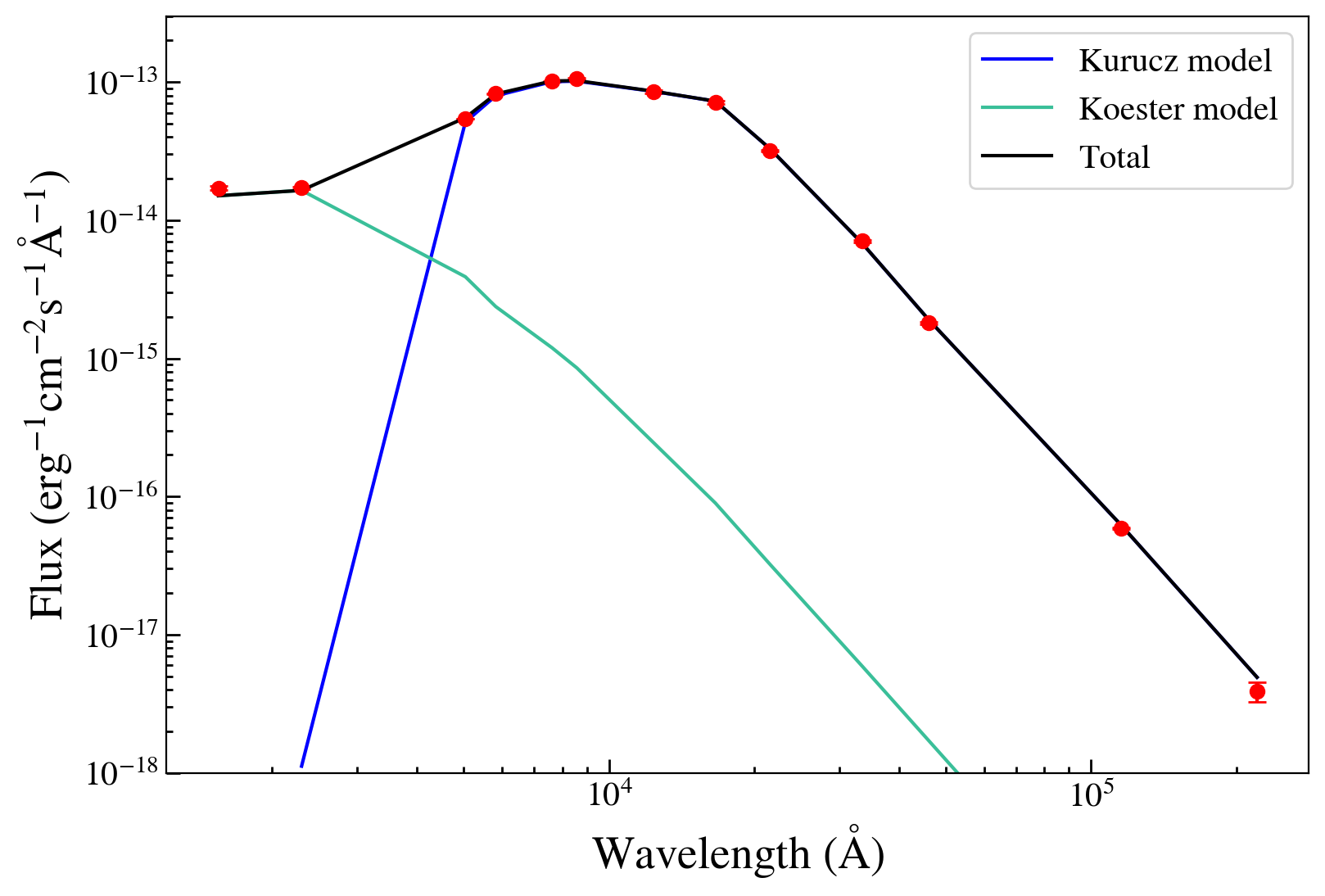}
	\caption{The best-fitting SED (black line) for GALEX J083531.3-090417 using GALEX, Gaia, 2MASS, and WISE broadband photometry (red dots). The blue line is the best Kurucz model fit for the red giant and the cyan line represent the best-fitting Koester model for the white dwarf. \label{vosa_sed_fit}}
\end{figure}

\begin{figure}[ht!] 
	\centering
        \includegraphics[width=0.45\textwidth]{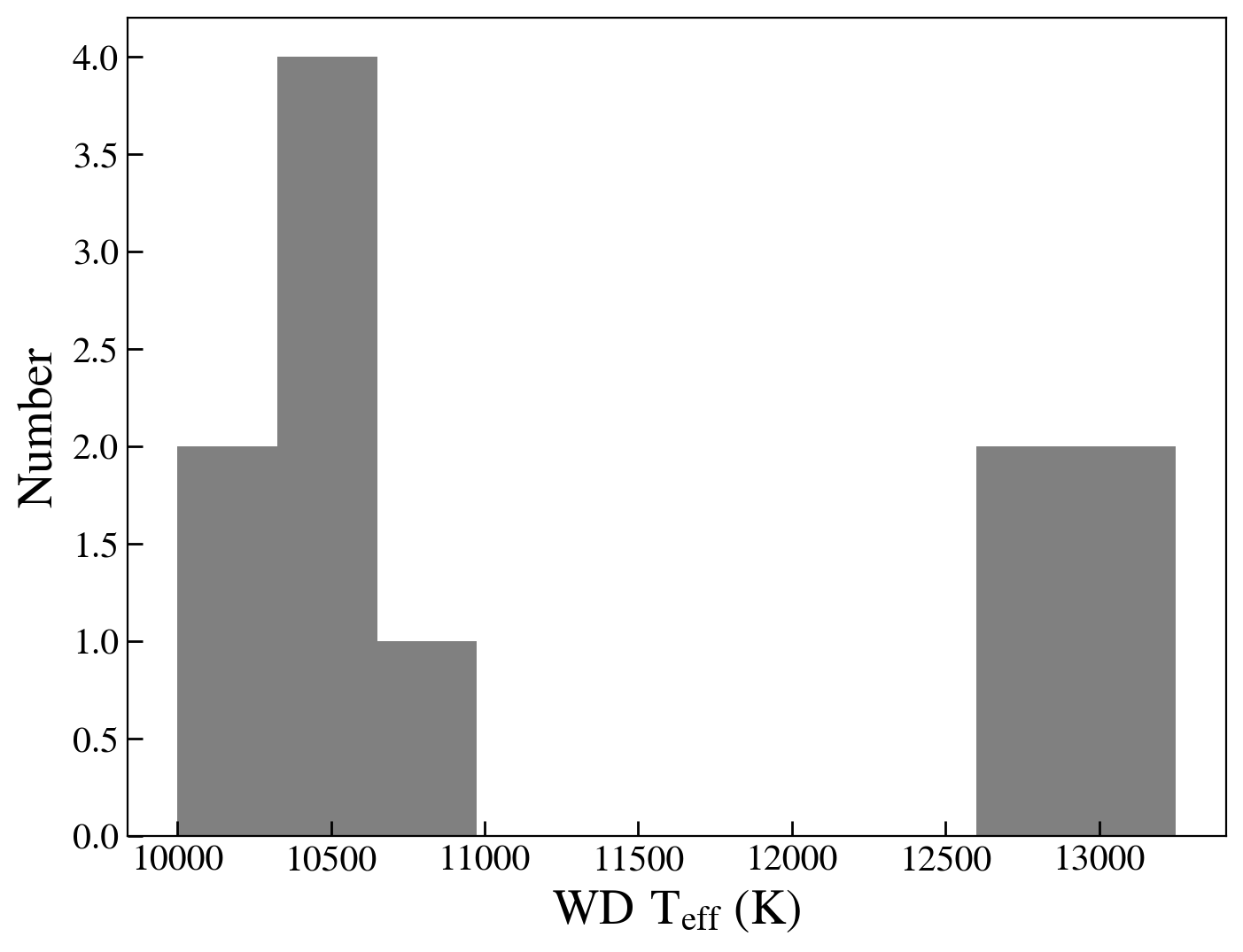}
	\caption{The white dwarf effective temperature distribution of the well-fitted 273 accreting-only symbiotic stars. \label{vosa_sed_wd_teff}}
\end{figure}

The $\chi^2$ fitting provides the best fit model and thus an estimation of the stellar parameters. Given that the effective temperature has a significant impact on the SED, while surface gravity and metallicity play relatively minor roles, we varied only the initial temperature parameter during the fitting process. Among the 12 acc-SySt candidates, satisfactory fits were achieved for 11 objects using a combination of white dwarf and Kurucz atmosphere models, while the remaining one candidate could not be well reproduced with these models. Figure \ref{vosa_sed_fit} presents a representative example of a well-fitted SED, and shows the best combined total fit (black line) to the observed fluxes (red dots). Figure \ref{vosa_sed_wd_teff} shows the temperature distribution of the 11 well-fitted sources. The best-fit temperature values and their associated uncertainties are tabulated in Table \ref{tab:Accret_inf}. In the process of VOSA binary fitting, the temperature uncertainties were conservatively estimated as half the parameter grid step, symmetrically centered on the best-fit value.

\subsection{The Accretion Rate} \label{sec:acc_rate}

The observed number of SySts is several orders of magnitudes lower than theoretical predictions, suggesting that the unobserved SySts may be with low accretion rates, which are insufficient to generate detectable emission lines. This could explain why they were not identified in previous surveys. To compare the accretion rates of acc-SySts and genetic SySts, we calculated their mass accretion rates using the following formula:
\begin{equation}
    M_{acc} = \frac{L_{acc}\cdot R_*}{G \cdot M_*},
\end{equation}
where $M_*$ and $R_*$ represent the mass and radius of the white dwarf, respectively, and $L_{acc}$ denotes the accretion luminosity. We can roughly estimate $L_{acc}$ using the formula as $L = 4 \pi d^2 f_x$, where $d$ is the distance and $f_x$ is the X-ray flux.

\begin{figure}[ht!] 
	\centering
        \includegraphics[width=0.45\textwidth]{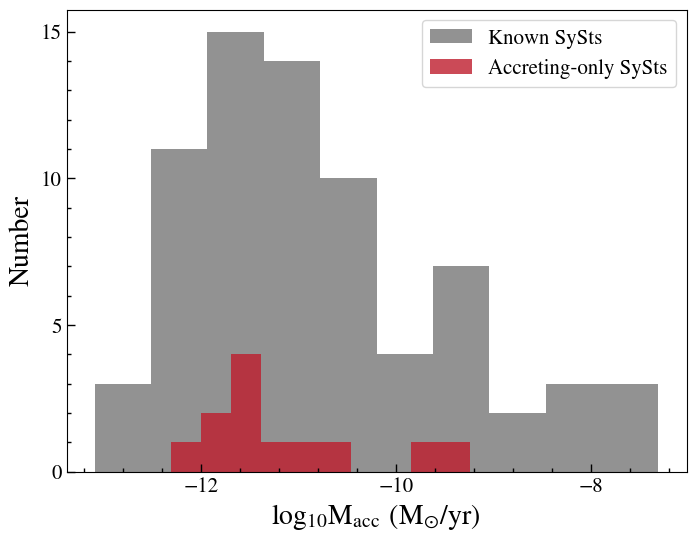}
	\caption{The mass accretion rate distribution of the acc-SySt candidates (red), and the known SySts (black), respectively. \label{Acc_rate}}
\end{figure}

Assuming a white dwarf mass of 0.8$M_{\odot}$, we estimated the accretion rate for the acc-SySt candidates, which were summarized in Table \ref{tab:Accret_inf}. The uncertainties in the accretion rates were calculated using the open-source Python package $uncertainties$ \footnote{\href{uncertainties}{https://uncertainties.readthedocs.io/en/latest/}}. For comparison, we cross-matched the known SySts from \cite{2019ApJS..240...21A} with the X-ray catalogs listed in Table \ref{tab:Xray_Catalog}, resulting in 73 common sources. We then estimated their accretion rates using the same method applied to the acc-SySt candidates. Figure \ref{Acc_rate} shows the  accretion rates distribution of acc-Syst candidates (red) and known SySts (black), respectively. Fig.\ref{Acc_rate} shows that the acc-SySts with X-ray fluxes exhibit accretion rates comparable to those of genuine SySts. \cite{1984ApJ...279..252K} demonstrated that the combination of RGB and WD with accretion rates below some limit would not be identified as SySts. This overlap suggests that a subset of acc-SySts could in fact be genuine SySts, however follow-up spectroscopic observations are required for confirmation.

\section{Summary} \label{sec:summary}

In this paper, we identified nine SySts from LAMOST DR10 database, comprising seven previously known and two newly discovered. Initially, we obtained and fitted the LAMOST DR10 pseudo-continuum spectrum, and selected spectra exhibiting typical emission lines characteristic of SySts, including $\rm H_{\alpha}, ~H_{\beta}, ~H_{\gamma}, ~and ~He II$. Subsequently, we filtered the emission spectra to identify SySt candidates based on magnitude and color criteria from the known SySts in HRD. Finally, we visually inspected these SySt candidates spectra and found nine SySts from LAMOST DR10 LRS, and classified these nine SySts as S-type SySts based on the $J -H$ vs. $H - Ks$ diagram.

Typical SySts with high accretion rates and strong emission lines are straightforwardly identified through spectroscopic observations. However, it is challenging to detect acc-SySts with lower accretion rates and either weak or undetectable emission lines using spectroscopic data alone. Therefore, we used GALEX photometric data to search for acc-SySt candidates. These candidates exhibited both UV and IR excess simultaneously. To minimize contamination, we cross-matched these candidates with the SIMABD database to exclude objects classified as Technetium-poor S stars, carbon stars, chemically peculiar stars, and SySts. Additionally, to ensure the presence of accretion activity, we cross-matched the acc-SySt candidates with several X-ray catalogs. Ultimately, we selected 12 acc-SySt candidates.

Furthermore, we estimated the temperatures of the white dwarfs in the acc-SySt candidates using the binary fitting tool available in VOSA. This method simultaneously fits the observed SEDs with Kurucz atmospheric models for the cool companion and Koester white dwarf models for the hot component. The white dwarf temperature corresponding to the model with the minimum Chi-square value was adopted as the best-fit temperature. The resulting white dwarf temperatures for these candidates lie in the range of 10,000--15,000 K. To investigate potential differences in accretion properties, we calculated the mass accretion rates for both the acc-SySt candidates and confirmed SySts. We found that the accretion rate distributions are remarkably similar between the two groups. This suggests that the acc-SySt candidates are likely genuine SySts.

\section{Acknowledgements}
This work is supported by the Postdoctoral Fellowship Program of CPSF under Grant Number GZC20232780. Y.B. Li is supported by the National Science Foundation of China (grant Nos. 12273078). Guoshoujing Telescope (the Large Sky Area Multi-Object Fiber Spectroscopic Telescope, LAMOST) is a National Major Scientific Project built by the Chinese Academy of Sciences. Funding for the Project has been provided by the National Development and Reform Commission. LAMOST is operated and managed by the National Astronomical Observatories, Chinese Academy of Sciences.
This research makes use of data from the European Space Agency (ESA) mission Gaia, processed by the Gaia Data Processing and Analysis Consortium.

This publication makes use of VOSA, developed under the Spanish Virtual Observatory (https://svo.cab.inta-csic.es) project funded by MCIN/AEI/10.13039/501100011033/ through grant PID2020-112949GB-I00.
VOSA has been partially updated by using funding from the European Union's Horizon 2020 Research and Innovation Programme, under Grant Agreement $\mathrm{n^\circ}$ 776403 (EXOPLANETS-A). This research also makes use of the VizieR \citep{2000AAS..143...23O} catalog access tool and the SIMBAD database  \citep{2000AAS..143....9W}, operated at Centre de Donnees astronomiques de Strasbourg (CDS), France.

\software~Astropy \citep{2013A&A...558A..33A}, TOPCAT \citep{2005ASPC..347...29T}, dustmaps \citep{2019ApJ...887...93G}, uncertainties \href{https://uncertainties.readthedocs.io/en/latest/}{https://uncertainties.readthedocs.io/en/latest/}.

$Facilities$: 2MASS, GALEX, WISE

\bibliography{symbiotic}{}
\bibliographystyle{aasjournal}

\end{document}